\documentstyle[12pt]{article}
\begin{document}
 \newcommand{\eqn}[1]{(\ref{#1})}
 \newcommand{\ft}[2]{{\textstyle\frac{#1}{#2}}}
 \newsavebox{\uuunit}
 \sbox{\uuunit}
     {\setlength{\unitlength}{0.825em}
      \begin{picture}(0.6,0.7)
         \thinlines
         \put(0,0){\line(1,0){0.5}}
         \put(0.15,0){\line(0,1){0.7}}
         \put(0.35,0){\line(0,1){0.8}}
        \multiput(0.3,0.8)(-0.04,-0.02){12}{\rule{0.5pt}{0.5pt}}
      \end {picture}}
 \newcommand {\unity}{\mathord{\!\usebox{\uuunit}}}
 \newcommand  {\Rbar} {{\mbox{\rm$\mbox{I}\!\mbox{R}$}}}
 \newcommand{\Ka}{K\"ahler}
 \def\ib{{\bar \imath}}
 \def\jb{{\bar \jmath}}
 \def\Im{{\rm Im ~}}
 \def\Re{{\rm Re ~}}
 \def\IP{\relax{\rm I\kern-.18em P}}
 \def\arccosh{{\rm arccosh ~}}

 \def\dop{{\rm d}\hskip -1pt}
 \def\bfone{\relax{\rm 1\kern-.35em 1}}
 \def\bfzero{\relax{\rm I\kern-.18em 0}}
 \def\inbar{\vrule height1.5ex width.4pt depth0pt}
 \def\IC{\relax\,\hbox{$\inbar\kern-.3em{\rm C}$}}
 \def\ID{\relax{\rm I\kern-.18em D}}
 \def\IF{\relax{\rm I\kern-.18em F}}
 \def\IH{\relax{\rm I\kern-.18em H}}
 \def\II{\relax{\rm I\kern-.17em I}}
 \def\IN{\relax{\rm I\kern-.18em N}}
 \def\IP{\relax{\rm I\kern-.18em P}}
\def\IK{\relax{\rm I\kern-.18em K}}
 \def\IQ{\relax\,\hbox{$\inbar\kern-.3em{\rm Q}$}}
 \def\IR{\relax{\rm I\kern-.18em R}}
 \def\IG{\relax\,\hbox{$\inbar\kern-.3em{\rm G}$}}
 \font\cmss=cmss10 \font\cmsss=cmss10 at 7pt
 \def\ZZ{\relax\ifmmode\mathchoice
 {\hbox{\cmss Z\kern-.4em Z}}{\hbox{\cmss Z\kern-.4em Z}}
 {\lower.9pt\hbox{\cmsss Z\kern-.4em Z}}
 {\lower1.2pt\hbox{\cmsss Z\kern-.4em Z}}\else{\cmss Z\kern-.4em
 Z}\fi}
 \def\a{\alpha} \def\b{\beta} \def\d{\delta}
 \def\e{\epsilon} \def\c{\gamma}
 \def\G{\Gamma} \def\l{\lambda}
 \def\L{\Lambda} \def\s{\sigma}
 \def\cA{{\cal A}} \def\cB{{\cal B}}
 \def\cC{{\cal C}} \def\cD{{\cal D}}
 \def\cF{{\cal F}} \def\cG{{\cal G}}
 \def\cH{{\cal H}} \def\cI{{\cal I}}
 \def\cJ{{\cal J}} \def\cK{{\cal K}}
 \def\cL{{\cal L}} \def\cM{{\cal M}}
 \def\cN{{\cal N}} \def\cO{{\cal O}}
 \def\cP{{\cal P}} \def\cQ{{\cal Q}}
 \def\cR{{\cal R}} \def\cV{{\cal V}}\def\cW{{\cal W}}
 %
 %
 %
 \def\crr{\crcr\noalign{\vskip {8.3333pt}}}
 \def\tilde{\widetilde}
 \def\bar{\overline}
 \def\us#1{\underline{#1}}
 \let\shat=\hat
 \def\hat{\widehat}
 \def\hyp{\vrule height 2.3pt width 2.5pt depth -1.5pt}
 \def\square{\mbox{.08}{.08}}
 \def\Coeff#1#2{{#1\over #2}}
 \def\Coe#1.#2.{{#1\over #2}}
 \def\coeff#1#2{\relax{\textstyle {#1 \over #2}}\displaystyle}
 \def\coe#1.#2.{\relax{\textstyle {#1 \over #2}}\displaystyle}
 \def\half{{1 \over 2}}
 \def\shalf{\relax{\textstyle {1 \over 2}}\displaystyle}
 \def\dag#1{#1\!\!\!/\,\,\,}
 \def\to{\rightarrow}
 \def\notin{\hbox{{$\in$}\kern-.51em\hbox{/}}}
 \def\shdot{\!\cdot\!}
 \def\ket#1{\,\big|\,#1\,\big>\,}
 \def\bra#1{\,\big<\,#1\,\big|\,}
 \def\equaltop#1{\mathrel{\mathop=^{#1}}}
 \def\Trbel#1{\mathop{{\rm Tr}}_{#1}}
 \def\inserteq#1{\noalign{\vskip-.2truecm\hbox{#1\hfil}
 \vskip-.2cm}}
 \def\attac#1{\Bigl\vert
 {\phantom{X}\atop{{\rm\scriptstyle #1}}\phantom{X}}}
 \def\exx#1{e^{{\displaystyle #1}}}
 \def\del{\partial}
 \def\delbar{\bar\partial}
 \def\nex#1{$N\!=\!#1$}
 \def\dex#1{$d\!=\!#1$}
 \def\cex#1{$c\!=\!#1$}
 \def\eg{{\it e.g.}} \def\ie{{\it i.e.}}
 %
 \def\cS{{\cal K}}
 \def\IE{\relax{{\rm I\kern-.18em E}}}
 \def\cE{{\cal E}}
 \def\rt{{\cR^{(3)}}}
 \def\IGam{\relax{{\rm I}\kern-.18em \Gamma}}
 \def\IGa{\IA}
 \def\LG{Lan\-dau-Ginz\-burg\ }
 \def\cV{{\cal V}}
 \def\Rt{{\cal R}^{(3)}}
 \def\wabc{W_{abc}}
 \def\WABC{W_{\a\b\c}}
 \def\W{{\cal W}}
 \def\tft#1{\langle\langle\,#1\,\rangle\rangle}
 \def\IA{\relax{\hbox{{\rm A}\kern-.82em {\rm A}}}}
 \let\picfuc=\fp
 \def\hata{{\shat\a}}
 \def\hatb{{\shat\b}}
 \def\hatA{{\shat A}}
 \def\hatB{{\shat B}}
 \def\bv{{Solv}}
 \def\spg{special geometry}
 \def\sc{SCFT}
 \def\leel{low energy effective Lagrangian}
 \def\pf{Picard--Fuchs}
 \def\pfS{Picard--Fuchs system}
 \def\el{effective Lagrangian}
 \def\Fb{\overline{F}}
 \def\nablab{\overline{\nabla}}
 \def\Ub{\overline{U}}
 \def\Db{\overline{D}}
 \def\zb{\overline{z}}
 \def\eb{\overline{e}}
 \def\fb{\overline{f}}
 \def\tb{\overline{t}}
 \def\Xb{\overline{X}}
 \def\Vb{\overline{V}}
 \def\Cb{\overline{C}}
 \def\Sb{\overline{S}}
 \def\delb{\overline{\del}}
 \def\Gammab{\overline{\Gamma}}
 \def\Ab{\overline{A}}
 \def\Anh{A^{\rm nh}}
 \def\alphab{\bar{\alpha}}
 \def\cy{Calabi--Yau}
 \def\cabg{C_{\alpha\beta\gamma}}
 \def\B{\Sigma}
 \def\Bh{\hat \Sigma}
 \def\Kh{\hat{K}}
 \def\Knh{{\cal K}}
 \def\A{\Lambda}
 \def\Ah{\hat \Lambda}
 \def\R{\hat{R}}
 \def\V{{V}}
 \def\T{T}
 \def\Gammah{\hat{\Gamma}}
 \def\twot{$(2,2)$}
 \def\K{K\"ahler}
 \def\rat{({\theta_2 \over \theta_1})}
 \def\lv{{\bf \omega}}
 \def\w{w}
 \def\CP{C\!P}
 \def\o#1#2{{{#1}\over{#2}}}
 \newcommand{\be}{\begin{equation}}
 \newcommand{\ee}{\end{equation}}
 \newcommand{\ba}{\begin{eqnarray}}
 \newcommand{\ea}{\end{eqnarray}}
 \newtheorem{definizione}{Definition}[section]
 \newcommand{\bd}{\begin{definizione}}
 \newcommand{\ed}{\end{definizione}}
 \newtheorem{teorema}{Theorem}[section]
 \newcommand{\bth}{\begin{teorema}}
 \newcommand{\eth}{\end{teorema}}
 \newtheorem{lemma}{Lemma}[section]
 \newcommand{\blem}{\begin{lemma}}
 \newcommand{\elem}{\end{lemma}}
 \newcommand{\brr}{\begin{array}}
 \newcommand{\err}{\end{array}}
 \newcommand{\nn}{\nonumber}
 \newtheorem{corollario}{Corollary}[section]
 \newcommand{\bcorol}{\begin{corollario}}
 \newcommand{\ecorol}{\end{corollario}}
 \def\twomat#1#2#3#4{\left(\begin{array}{cc}
  {#1}&{#2}\\ {#3}&{#4}\\
 \end{array}
 \right)}
 \def\twovec#1#2{\left(\begin{array}{c}
 {#1}\\ {#2}\\
 \end{array}
 \right)}
 \begin{titlepage}
\hskip 8cm\vbox{\hbox{CERN TH/96-315}
\hbox{November 1996}}
\vskip 1cm
 \begin{center}
{\large {R--R Scalars, U--Duality
   and
   Solvable Lie Algebras\footnote{Work supported in part
   by EEC under TMR contract ERBFMRX-CT96-0045
and by DOE grant DE-FGO3-91ER40662
 } }
  }
  \vskip 1cm
{\bf L. Andrianopoli$^{1}$, R. D'Auria$^{2}$,
 S. Ferrara$^{3}$,\\
 P.Fr\'e$^{4}$  and M. Trigiante$^{5}$ \\}
\vskip 0.5cm
{\small  $^1$ Dip. Fisica, Universit\'a di Genova,
Via Dodecaneso 33, I-16146 Genova\\
and INFN - Sez. Genova, Italy\\
$^2$ Th. Phys. Division CERN, CH 1211 Geneva 23, Switzerland
\footnote{on leave from Politecnico di Torino C.so Duca
degli Abruzzi 24, I-10129 Torino}\\
and INFN - Sez. di Torino, Italy\\
$^3$ Th. Phys. Division CERN, CH 1211 Geneva 23, Switzerland \\
and  INFN, L.N.F., Italy
 \\
$^4$  Dip. Fisica Teorica, Universit\`a di Torino,
  Via P. Giuria 1, I-10125 TORINO, Italy and INFN, Sez. Torino \\
$^5$ SISSA, Via Beirut 4, I-34100 Trieste, Italy and INFN Sez. Trieste }
\end{center}
\vskip 0.2cm
 \begin{center}
{\bf Abstract}
\end{center}
 We consider the group theoretical properties of R--R scalars
 of string theories
in the low--energy supergravity limit and relate them to the solvable
Lie subalgebra $\IG_s\subset {\bf U}$ of the U--duality algebra
that generates the scalar manifold of the theory:
$\exp[\IG_s]= U/H$.
Peccei--Quinn symmetries are naturally related with the maximal
abelian ideal ${\cal A} \subset \IG_s $ of the solvable Lie algebra.
The solvable algebras of maximal rank  occurring in maximal
supergravities in diverse
dimensions are described in some detail.
A particular  example of a solvable Lie algebra is a rank one,
$2(h_{2,1}+2)$--dimensional
 algebra displayed by the classical quaternionic spaces that
are obtained via c--map from the special K\"ahlerian moduli
spaces of Calabi--Yau threefolds.

\end{titlepage}

\section{Introduction}
One of the most intriguing features of string theory, in its perturbative
formulation, is the existence of two kind of scalars, those coming
from the Neveu--Schwarz sector (N--S) and those coming from the
Ramond--Ramond sector (R--R). The former fields have sometimes
the interpretation of moduli of a conformal field--theory, while
the latter  do not have such a property \cite{gsw}.
 However,
in an effective lagrangian formulation for the string light--states,
the R--R scalars are linked to N--S scalars by supersymmetries
exchanging left and right movers and more interestingly by
U--dualities \cite{ht}, which, for continuous transformations, are
related to the non--compact symmetries present in extended
supergravities \cite{csf}, \cite{cj}. These symmetries,
not present in the perturbative string
spectrum, are conjectured to be symmetries at the non--perturbative level,
at least under the restriction $U \, \to \, U(\ZZ)$.
Indeed supergravity theories in  diverse dimensions  \cite{sase}
constitute a nested web filling a plane whose axes are
the space--time dimensions $D$ and the number of supersymmetry
charges $N$. Our recently improved
understanding  of  non perturbative   string
theory  has taught us to regard all the
lagrangians in the web as different effective actions describing
the interaction of the light fields in
different corners of a single {\it quantum theory}. The glue
that keeps the various parts of the web together is provided by
duality transformations \cite{rev}. Although the ideas are
conceptually new
their mathematical realization occurs by means of structures that
have been known for many years. Indeed the relevant duality
transformation groups  are  nothing else but the well known
hidden symmetries of supergravity  governing the structure
of the scalar sector \cite{cre}. In every dimension $D$ and for each
value of $N$ the $n_s$ scalar fields $\varphi^I$
can be interpreted, at least locally, as the coordinates of an appropriate
Riemanian manifold
${\cal M}_{scalar}$ whose metric $g_{IJ}(\varphi)$ appears in the
scalar kinetic term
\begin{equation}
{\cal L}^{scal}_{kin} \, = \,
\o{1}{2} \, g_{IJ}(\varphi) \, \partial^\mu \varphi^I \,
 \partial_\mu \varphi^J
\end{equation}
Let $U$ be the group of isometries (if any) of the scalar
metric $g_{IJ}(\varphi)$. The elements of $U$ correspond  to  global
symmetries  of the $\sigma$--model lagrangian
${\cal L}^{scal}_{kin}$. If the action of $U$
 were not extended from the scalar fields also to
the other fields and in particular to the vector fields or
higher rank $p$--form potentials, $U$ could not be
 promoted to a symmetry of the full theory.
 This is clear from the fact that
the scalar fields are related to the spinor and vector fields,
and/or ($p+1$)--form potentials, by supersymmetry
 transformations. The implementation of
 $U$ isometries  in a supersymmetric
consistent way is the basic issue of supergravity hidden symmetries.
Indeed a generic feature of supergravity theories
at $D=4$ is that they have a symmetry under ``duality''
which acts non linearly on the scalars and linearly on the
 field-strengths and their duals, that are fitted together into
 a single suitable symplectic representation of $U$ \cite{gz}.

In diverse dimensions $D
$ the U--duality group acts linearly on generic ($p+1$)--potentials
unless $p+2={D\over 2}$
in which case it acts linearly on the field strengths and their
duals \cite{sase}.
\par
The low energy supergravities divide into two classes:
\begin{enumerate}
\item{the first, containing  the $D=4$, $N\le 2$ and the $D=5$,
$N=2$ cases,  is the class where the scalar manifold
 $\cM_{scalar}$ can admit isometries, but it is not necessarily
 a homogeneous space $U/H$}
\item{All the other theories have scalar manifolds which are
necessarily homogeneus coset
manifolds $U/H$ and this class comprises, in particular,
the $D=4$, $N>2$ theories
 and all the maximally extended supergravities
in $D\le 11$
}
\end{enumerate}
In the first class of theories the local scalar geometries
defined at string tree level acquire perturbative
 and non--perturbative quantum corrections.
In the second class, which will be the main focus of this paper,
 the local scalar geometry given by the natural Riemannian
metric defined on $U/H$ is protected by supersymmetry
against quantum corrections.
\par
In this paper we  investigate
the solvable algebra $Solv(U/H)=\IG_s\subset U$ that generates
the Riemannian manifold $U/H$
 in such a way that $\exp [\IG_s]=U/H$
 and hence $\mbox{dim}\,
\IG_s=\mbox{dim}\,U -\mbox{dim}\,H$.
\par
Solvable Lie algebras appeared first in the supergravity
literature to classify
quaternionic manifolds with a transitive, solvable group
of motions \cite{ale}, \cite{bw}\cite{cec}
\cite{dwvp}\cite{cfg}.

The construction of $\IG_s$ is available in standard textbooks
\cite{hel}.
It suffices to say that $\IG_s$ contains a non compact Cartan
 part $\cH_K$ which is the abelian set of semisimple generators
 in $\IK$ defined by the Cartan decomposition
\begin{equation}
{\bf U}=\IH \oplus \IK
\end{equation}
and other nilpotent  generators coming from the positive roots
of ${\bf U}$.\\
Among them, of particular relevance is the maximal abelian
ideal $\cA$ whose elements have the
physical interpretation of being the translational
(Peccei--Quinn) isometries of the theory.

The use of the solvable Lie algebra representation allows
one to regard the coset
manifold $U/H$ as the group manifold of the corresponding
solvable group with its own advantages.
For instance all
 the geometric notions (metric, connection, curvature)
 are translated into an
 algebraic language and, in particular, one obtains an
 intrinsic privileged set
 of coordinates for the manifold where each scalar field
 is in one to one correspondence
 with a generator of the solvable algebra.
\par
The natural question which arises is therefore that of
finding an intrinsic algebraic
 characterization of the R--R scalars (i.e. R--R generators)
 with respect to
the N--S scalars (i.e. N--S generators), which is otherwise
obscure in the effective supergravity formalism.
\par
In this paper we show how to obtain this characterization by
decomposing the
$U$--duality algebra, and hence its solvable subalgebra, with respect to
its S--duality and T--duality subalgebras. Indeed the distinction between
Neveu--Schwarz (N--S) and Ramond (R--R) scalars is T--duality invariant.
\par
Mastering the structure of the solvable Lie algebra appears
to be relevant in different respects.
 A particularly significant one is partial supersymmetry
breaking. It appears from recent results \cite{fgp}, \cite{fgpt},
obtained in the context
 of  N=2 theories,  that partial SUSY breaking $N=2 \to N=1$,
 with zero vacuum energy,
 can be obtained precisely by gauging generators in the maximal abelian
ideal $\cA\subset \IG_s$ of the solvable algebra \cite{fgpt}.
In this respect  fields of the maximal abelian ideal
contain the flat directions after
gauging.
\par
We expect the same to be true in other extended theories with
the eigenvalues
 of the gravitino mass matrix  parametrized by the charges of the fields
with respect to $\cA$. It goes without saying that, whenever these
charges are of R--R type \cite{post} they carry a
non--perturbative significance so that
knowledge of $\cA$ and of its N--S, R--R splitting
is a fundamental prerequisite.
This is the information we present in this paper.

It is hoped that some of the properties outlined in the
present investigation
 may also be useful to explore features of R--R scalars
 when the moduli space gets quantum corrected.
This is known to occur, through D--two branes instanton
effects in type IIA theory \cite{bbs} compactified on
Calabi--Yau manifolds, as a consequence of second quantized
mirror symmetry \cite{fhsv}.
In particular such corrections should resolve conifold
singularities \cite{s gms}, \cite{oova}, \cite{sw s ss}
in the construction of quaternionic manifolds by $c$--map \cite{cfg}
 from the complex structure moduli space of Calabi--Yau
threefolds.

\section{Solvable Lie Algebras: the machinery.}
In this section we will deal with a general property according to which
any homogeneous non-compact coset manifold may be expressed as
a group manifold
 generated by a suitable solvable Lie algebra. \cite{ale}
\par
Let us start by giving few preliminar definitions.
A {\it solvable } Lie algebra $\IG_s$ is a Lie algebra whose $n^{th}$ order
(for some $n\geq 1$) derivative algebra vanishes:
\begin{eqnarray}
{\cal D}^{(n)}\IG_{s}&=&0 \nonumber \\
\cD\IG_s=[\IG_s,\IG_s]&;&\quad \cD^{(k+1)}\IG_s=[\cD^{(k)}\IG_s,
\cD^{(k)}\IG_s]\nonumber
\end{eqnarray}
A {\it metric} Lie algebra $(\IG,h)$ is a Lie algebra endowed with an
euclidean metric $h$. An important theorem states that if a Riemannian
manifold
$(\cM,g)$ admits a transitive group of isometries $\cG_s$ generated by
a solvable Lie algebra $\IG_s$ of the same dimension as $\cM$, then:
\begin{eqnarray}
\cM\sim \cG_s&=&exp(\IG_s)\nonumber\\
 g_{|e\in \cM}&=&h \nonumber
\end{eqnarray}
where $h$ is an euclidean metric defined on $\IG_s$.
 Therefore there is a one to one correspondence between
 Riemannian manifolds fulfilling the hypothesis
stated above and solvable metric Lie algebras $(\IG_s,h)$.\\
Consider now an homogeneous coset manifold $\cM=\cG /\cH$,
$\cG$ being a non compact real
 form of a semisimple Lie group and $\cH$ its maximal compact
subgroup. If $\IG$ is the Lie algebra generating $\cG$, the so
called Iwasawa
decomposition ensures the existence of a solvable Lie subalgebra
$\IG_s\subset
\IG$, acting transitively on $\cM$, such that \cite{hel}:
\begin{equation}
\IG=\IH\oplus \IG_s \qquad \mbox{dim }\IG_s=\mbox{dim }\cM \nonumber
\end{equation}
$\IH$ being the maximal compact subalgebra of $\IG$ generating $\cH$. \\
In virtue of the previously stated theorem, $\cM$ may be expressed
 as a solvable
 group manifold generated by $\IG_s$. The algebra $\IG_s$ is constructed
as follows \cite{hel}. Consider the Cartan decomposition
\begin{equation}
\IG = \IH \oplus \IK
\end{equation}
Let us denote by $\cH_K$ the  maximal abelian subspace
of $\IK$ and by $\cH$
the Cartan subalgebra of $\IG$.
It can be proven \cite{hel} that $\cH_K = \cH \cap \IK$,
that is it consists of all non compact elements of $\cH$.
Furthermore let $h_{\alpha_i}$ denote the elements
of $\cH_K$, $\{\alpha_i\}$ being
 a subset of the positive roots of $\IG$ and $\Phi^+$
 the set of positive roots $\beta$ not orthogonal to
all the $\alpha_i$ (i.e. the corresponding ``shift'' operators
$E_\beta$ do not commute with $\cH_K$).
It can be demonstrated that the solvable algebra $\IG_s$
defined by the Iwasawa decomposition
may be expressed in the following way:
\begin{equation}
  \label{iwa}
  \IG_s = \cH_K \oplus \{\sum_{\alpha \in \Phi^+}E_\alpha \cap \IG \}
\end{equation}
where  the intersection with $\IG $ means that $\IG_s$ is generated
by those suitable complex combinations of the ``shift'' operators
which belong to the real form of the isometry algebra $\IG$.
\par
The {\it rank} of an homogeneous  coset manifold is defined as
the maximum number of commuting semisimple
elements of the non compact subspace $\IK$. Therefore it
coincides with the dimension of $\cH_K$,
i.e. the number of non compact Cartan generators of $\IG$.
A  coset manifold is {\it maximally non compact} if
$\cH =\cH_K \subset \IG_s$.
The relevance of maximally non compact coset manifolds relies
on the fact that they are spanned by
the scalar fields in the maximally extended supergravity theories.
\par
As an example of the procedure just described we will work out
the solvable Lie algebra
corresponding to the manifold \cite{dwvp}
\begin{equation}
  \cM = {SU(1,n+2)\over U(1) \otimes SU(n+2)}
\end{equation}
whose rank is one. Indeed, if we express the roots $\alpha$
of  $SU(1,n+2)$
as $\epsilon_i -\epsilon_j$, $1\le i<j\le n+3$, the only non
compact element of the Cartan subalgebra of $SU(1,n+2)$
is $H_{\epsilon_1 -\epsilon_{n+3}}$, ($H_\alpha, E_\alpha$)
denoting the canonical basis
of the $SU(1,n+2)$ algebra.
The positive roots of $\Phi^+$ are the ($2n+3$) of the form
$\epsilon_1 -\epsilon_i, \epsilon_j
 -\epsilon_{(n+3)}$  ($i=2,\cdots,n+3$, $j=2,\cdots,n+2$).
According to (\ref{iwa}), the generators of $\IG_s$ are:
\begin{equation}
  \{H_{\epsilon_1 -\epsilon_{n+3}},
  (E_{\epsilon_1 -\epsilon_i},E_{\epsilon_j
 -\epsilon_{(n+3)}})\cap SU(1,n+2)\}
\end{equation}
Defining $x_{i}=E_{\epsilon_1 -\epsilon_i}$, $y^i= E_{\epsilon_i
 -\epsilon_{(n+3)}}$, $i=2,\cdots, n+2$,
 $h=H_{\epsilon_1 -\epsilon_{n+3}}/2$
and $z=E_{\epsilon_1 -\epsilon_{n+3}}$, one can check
that these generators fulfill
the commutation relations in equations (\ref{heisen}),  (\ref{cseral})
 of next section, which  characterize the action of the
 isometries of a special quaternionic manifold on the R--R scalars.

\section{$c$--map, special quaternionic manifolds and their
solvable Lie algebra}
The simplest example of solvable Lie algebra occuring in an
effective supergravity theory is found while considering the
$c$--map \cite{cfg} of special K\"ahler manifolds in the context of string
compactifications on Calabi--Yau threefolds. In this context
the classical quaternionic geometry of $N=2$ hypermultiplets
for type II strings is given by {\it special quaternionic manifolds}
${\cal SQM}$ of real dimension $\mbox{dim}_{\bf R}\, {\cal SQM}$=$
4 \, h_{(2,1)} + 4$ of which half are Ramond--Ramond scalars
$C_\Lambda$, ($\Lambda = 0,1\,\cdots\,h_{(2,1)}$)  and
the other half are Neveu--Schwarz scalars.
The latter include
the axion--dilaton degree of freedom ${\cal S}$ and the
$h_{(2,1)}$ moduli $z^{i}$ ($i = 1 \,\cdots\,h_{(2,1)}$)
of the Calabi--Yau threefold complex structures.
\par
In \cite{fesa} it was observed that a generic special quaternionic
metric has always a $4 + 2 h_{(2,1)}$ dimensional group of isometries
which act on the R--R (complex) scalars $C_\Lambda$
and the axion--dilaton system  S as follows:
\begin{eqnarray}
{\cal S}^\prime & = & {\cal S} +{\rm i} \alpha - 2\,
C_{\Lambda} \, \gamma^\Lambda - \gamma^\Lambda \,
{\cal N}_{\Lambda\Sigma} \, \gamma^\Sigma \nonumber\\
C^{\prime}_\Lambda & =& C_\Lambda + {\rm i} \beta_\Lambda
+ {\cal N}_{\Lambda\Sigma} \, \gamma^\Sigma \nonumber\\
{\cal S}^\prime & = & \lambda \, {\cal S} \nonumber\\
C^{\prime}_\Lambda & =&\lambda^{1/2} \,  C_\Lambda
\label{cmappa}
\end{eqnarray}
where ${\alpha}, \lambda$, ${\gamma^\Lambda} ,
\beta_\Lambda$, ($\Lambda=0,\cdots , h_{(2,1)}$) are real parameters
and $\cN_{\Lambda\Sigma}$ is a symmetric
matrix depending on the moduli $z^i, \bar z^i$.
For infinitesimal ${\alpha}, \lambda$, ${\gamma^\Lambda} ,
\beta_\Lambda$ transformations the corresponding generators
$z, h,$ $y^\Lambda , x_\Lambda$ satisfy the following Lie algebra:
\begin{eqnarray}
\left [ x_\Lambda \, , \, y^\Sigma \right ] & = &
\delta^\Sigma_\Lambda \, z \nonumber\\
\left [ x_\Lambda \, , z \right ] &= &
\left [ y^\Lambda \, , z \right ] \, = \,
\left [ x_\Lambda \, , \, x_\Sigma \right ] \, = \,
\left [ y^\Lambda \, , \, y^\Sigma \right ]\, = \, 0
\label{heisen}
\end{eqnarray}
\begin{equation}
\left [ h \, , \, z \right ]= z ;\quad
\left [ h \, , \, x_\Sigma \right ] = {\o{1}{2}}
\,  x_\Sigma ;\quad
\left [ h \, , \, y^\Sigma \right ] = {\o{1}{2}}
\,  y^\Sigma
\label{cseral}
\end{equation}
where eq. \ref{heisen} define a $2 \, h_{(2,1)} +3$
nilpotent Lie algebra.
When extended with the $h$  generator it becomes a
$2 \, h_{(2,1)} +4$ solvable Lie algebra with a
one--dimensional Cartan subalgebra ${\cal H}_S = h $.
\par
The  Lie algebra in eq.s (\ref{heisen}), (\ref{cseral}) is nothing else
but the solvable Lie algebra
${Solv} \,\left ( {\cal F}_{h_{(2,1)}} \right )$
generating the coset
\begin{equation}
{\cal F}_{h_{(2,1)}}\, \equiv \,
 {\o {SU(1,h_{(2,1)}+2)}{U(1)\otimes
SU(h_{(2,1)}+2)}}
\end{equation}
This is is simply a consequence of the fact that the
special quaternionic manifolds can be viewed as a
${\cal F}_{h_{(2,1)}}$--fibration over the $h_{(2,1)}$ dimensional Special
K\"ahlerian moduli space.  In other words, the fiber above
each point in moduli space is diffeomorphic and isometric
to ${\cal F}_{h_{(2,1)}}$. This is the pointwise
splitting into the special K\"ahler base manifold
and the R--R + axion--dilaton fiber.
 The maximal abelian ideal of
${Solv} \,\left ( {\cal F}_{h_{(2,1)}} \right )$ has therefore
dimension $h_{(2,1)}+2$ of which $h_{(2,1)}+1$ are
Ramond generators and $1$ is a Neveu--Schwarz generator.
\section{Maximal rank solvable Lie algebras:
N--S and R--R scalars for maximal
supergravities in diverse dimensions}
Let us consider the list of maximally extended supergravities
that are obtained
dimensionally reducing $D=11$ supergravity \cite{cjs} on a ($11-D$)--torus,
and keeping all the massless modes. In this case the U-duality algebra is
$E_{11-D(11-D)}$ \cite{cre}, namely   that real section of
the complex Lie algebra $E_{11-D}$ which is maximally non-compact.
To explain the notations:
 by $E_{n(r)}$ we denote the
real form of the rank $n$ complex Lie algebra $E_{n}$,
where $r\le n$ Cartan generators are non--compact: when $r=n$
all Cartan generators are non compact and from section 2 we know that this
 is the case where the total number of non--compact generators is
maximum. Indeed, when $\cH_K= \cH$ all the positive  roots are included in
the solvable Lie algebra. This latter has  therefore the
universal simple form:
\begin{equation}
\IG_s \, = \, {\cal H} \, \oplus \,
\sum_{\alpha \, \in \, \Phi^+}  \, E^{\alpha}
\label{simstru}
\end{equation}
where ${\cal H}$ is the Cartan subalgebra, $E^{\alpha}$ is
the root--space
corresponding to the root $\alpha$ and $\Phi^+$
denotes the set of
positive roots of the U--duality group ($E_{11-D(11-D)}$).
 The scalar fields  parametrize the coset manifold $E_{11-D(11-D)}/H$
where $H$ is the maximal compact subgroup $H \subset E_{11-D(11-D)}$.
 The number $r=11-D$, which is the rank of both the
U--duality algebra and of the scalar manifold, is
by its own definition  the number of compactified dimensions.
\par
In fact  the Cartan semisimple piece $\cH=O(1,1)^{11-D}$ of the
solvable Lie algebra
has the physical meaning of
\footnote{Similar reasonings appear in refs.\cite{maxsg}\cite{stellelu}}
  diagonal moduli for the $T_{11-D}$
compactification torus
(roughly speaking the radii of the $11-D$ circles)
(in modern language, this is the M--theory interpretation)
\cite{wit}.
\par
From  a stringy (type IIA) perspective one of them is the dilaton
and the others are the Cartan piece of the maximal rank solvable
Lie algebra  generating the moduli space ${O(10-D,10-D)\over O(10-D)
\otimes O(10-D)}$  of the $T_{10-D}$ torus.
\par
This trivially implies that the Cartan piece is always
in the N--S sector.
\par
We are interested in splitting the maximal solvable subalgebra
(\ref{simstru}) into its N--S and R--R parts.
To obtain this splitting, as already mentioned in the introduction,
we just have to decompose
the U--duality algebra $U$ with respect to its ST--duality
subalgebra $ST\subset U$ \cite{feko}, \cite{wit}.\footnote{
Note that at $D=3$, ST--duality merge in a simple Lie algebra
\cite{sm}\cite{sen}.}
We have:
\begin{eqnarray}
  5 \leq D \leq 9 \quad :\quad & ST = &O(1,1)
  \otimes O(10-D,10-D) \nonumber\\
 D = 4 \quad :\quad & ST = &Sl(2,\IR) \otimes O(6,6) \nonumber\\
 D =3 \quad :\quad & ST = & O(8,8)
\label{stdual}
\end{eqnarray}
Correspondingly we obtain the decomposition:
\begin{eqnarray}
 5 \leq D \leq 9 \quad :\quad  \mbox{adj} \,
 E_{11-D(11-D)}&= & \mbox{adj}O(1,1) \oplus \mbox{adj}O(10-D,10-D)
 \nonumber\\
&&\oplus \left(2,\mbox{spin}_{(10-D,10-D)} \right)\nonumber\\
 D=4 \quad :\quad  \mbox{adj} \, E_{7(7)}&= & \mbox{adj}Sl(2,\IR)
 \oplus \mbox{adj}O(6,6)
\oplus \left(2,\mbox{spin}_{(6,6)} \right)\nonumber\\
 D=3 \quad :\quad  \mbox{adj} \, E_{8(8)}& = &  \mbox{adj}O(8,8)
\oplus \mbox{spin}_{(8,8)}
 \label{stdec}
 \end{eqnarray}
From (\ref{stdec}) it follows that:
\begin{eqnarray}
  \label{stdim}
  5 \leq D \le 9 \quad &:& \quad  \mbox{dim} E_{11-D(11-D)}= 1 +
  (10-D)(19-2D) + 2^{(10-D)}  \nonumber\\
D= 4  \quad &:& \quad  \mbox{dim}E_{7(7)}=
\mbox{dim}[ \left ( {\bf 66},{\bf 1} \right )
\oplus \left ( {\bf 1},{\bf 3} \right ) \oplus \left ({\bf 2},{\bf 32}
\right )]\nonumber\\
D=3 \quad &:& \quad \mbox{dim}E_{8(8)}=  \mbox{dim}[ {\bf 120}
\oplus {\bf 128}]
\end{eqnarray}
The dimensions of the maximal rank solvable algebras are instead:
\begin{eqnarray}
  \label{solvdim}
  5 \le D\le 9 \quad &:& \quad  \mbox{dim} \IG_s=  (10-D)^2 +  1 +
  2^{(9-D)} = \mbox{dim} {U\over H} \nonumber\\
D= 4 \quad &:& \quad   \mbox{dim}\IG_s =    32 +  37 + 1 = \mbox{dim}
{U\over H} \nonumber\\
D = 3 \quad &:& \quad \mbox{dim}\IG_s=  64 +  64  = \mbox{dim} {U\over H}
\end{eqnarray}
The above parametrizations of the dimensions of the cosets
listed in Table 1
can be traced back to the fact that the N--S and R--R
generators are given respectively by:
\begin{equation}
  \mbox{ N--S} = \mbox{Cartan generators} \oplus
  \mbox{positive roots of }\mbox{adj}\,ST
\end{equation}
and
\begin{equation}
   \mbox{R--R}= \mbox{positive weights of }\mbox{spin}_{ST}
\end{equation}
In this way we have:
\begin{eqnarray}
 \mbox{dim}(\mbox{ N--S})&=&\cases{ (10-D)^2 + 1 \quad\quad (5 \le D\le 9)
 \cr
  38= 7+1+30  \quad\quad (D=4)\cr
64=8+56 \quad\quad (D=3)\cr}\nonumber\\
\mbox{dim}(\mbox{ R--R})&=&\cases{ 2^{(9-D)} \quad\quad
(5 \le D \le 9) \cr
 32  \quad\quad(D=4)\cr
64 \quad\quad (D=3)\cr}
\end{eqnarray}
\par

For $D=3$ we notice that the ST--duality group $O(8,8)$
is a non compact form of the $\IH$
maximal compact subgroup $O(16)$ of the U--duality group $E_{8(8)}$.
\par
This explains why R--R = N--S = 64 in this case.
Indeed, 64 are the positive weigths of
$\mbox{dim}(\mbox{spin}_{16}) = 128$.
This coincides with the counting of the bosons
in the Clifford algebra of $N=16$
supersymmetry at $D=3$.

\par
\begin{table}[h]
\begin{center}
\begin{tabular}{|c|c|c|c|}
\hline
$D=9$ &      $E_{2(2)}  \equiv SL(2,\IR)\otimes O(1,1)$ & $H =
O(2) $ &
$\mbox{dim}_{\bf R}\,(U/H) \, =\, 3$ \\
 \hline
$D=8$ &      $E_{3(3)}  \equiv SL(3,\IR)\otimes Sl(2,\IR)$ & $H =
O(2)\otimes O(3) $ &
$\mbox{dim}_{\bf R}\,(U/H) \, =\, 7$ \\
\hline
$D=7$ &      $E_{4(4)}  \equiv SL(5,\IR) $ & $H = O(5) $ &
$\mbox{dim}_{\bf R}\,(U/H) \, =\, 14$ \\
\hline
$D=6$ &      $E_{5(5)}  \equiv O(5,5) $ & $H = O(5)\otimes O(5) $ &
$\mbox{dim}_{\bf R}\,(U/H) \, =\, 25$ \\
\hline
$D=5$ &      $E_{6(6)}$   & $H = Usp(8) $ &
$\mbox{dim}_{\bf R}\,(U/H) \, =\, 42$ \\
\hline
$D=4$ &      $E_{7(7)}$   & $H = SU(8) $ &
$\mbox{dim}_{\bf R}\,(U/H) \, =\, 70$ \\
\hline
$D=3$ &      $E_{8(8)}$   & $H = O(16) $ &
$\mbox{dim}_{\bf R}\,(U/H) \, =\, 128$ \\
\hline
\end{tabular}
\caption{U--duality groups and maximal compact subgroups of maximally
extended supergravities.}
\label{costab}
\end{center}
\end{table}
In  Table 2 we give, for each of the previously listed cases,
 the dimension of the maximal abelian ideal $\cA$ of the solvable
 algebra and its
N--S, R--R content
\cite{cre}, \cite{ht}.
\begin{table}[h]
  \begin{center}
    \begin{tabular}{|c|c|c|c|}
\hline
D  & dim $\cA$ & N--S & R--R \\
\hline
3 & 36 & 14 & 22 \\
\hline
4 & 27 & 11 & 16 \\
\hline
5 & 16 & 8 & 8 \\
\hline
6 & 10 & 6 & 4 \\
\hline
7 & 6 & 4 & 2 \\
\hline
8 & 3 & 2 & 1 \\
\hline
9 & 1 & 0 & 1 \\
\hline
    \end{tabular}
    \caption{Maximal abelian ideals.}
    \label{tab:2}
  \end{center}
\end{table}


\section{Electric subgroups}
In view of possible applications to the gauging
of isometries of the four dimensional U--duality group, which may
give rise to spontaneous partial supersymmetry breaking
with zero--vacuum energy \cite{fgp}, \cite{fgpt}, it is relevant
to answer the following question: what is
the electric subgroup\footnote{By ``electric''
we mean the group which has a lower triangular
symplectic embedding, i.e. is a symmetry of the
lagrangian \cite{dwvp1}, \cite{cdfvp}.}
of the solvable group? Furthermore, how many of
 its generators are
of N--S type and how many are of R--R type?
Here as an example we focus on the maximal
$N=8$ supergravity in $D=4$.
To solve the problem we have posed
we need to consider the splitting of the
U--duality symplectic representation pertaining
to vector fields, namely the  ${\bf 56}$ of $E_{7(7)}$,
under reduction with respect to the ST--duality
subgroup.
 The fundamental ${\bf 56}$ representation
defines the symplectic embedding:
\begin{equation}
E_{7(7)} \, \longrightarrow \, Sp(56,\IR)
\label{embed}
\end{equation}
We have:
\begin{equation}
{\bf 56} \, {\stackrel{Sl(2,R)\otimes SO(6,6)}
{\longrightarrow}} \, \left ({\bf 2} ,
{\bf 12}\right ) \, \oplus \, ({\bf 1},{\bf 32})
\end{equation}
This decomposition is understood from the physical
point of view by noticing that
the $28$ vector fields split into $12$ N--S fields
which, together with their
magnetic counterparts, constitute the $\left ({\bf 2} ,
{\bf 12}\right )$ representation plus $16$
R--R fields whose electric and magnetic
field strenghts build up the {\it irreducible} ${\bf 32}$ spinor
 representation of
$O(6,6)$. From this it follows that the T--duality
 group is purely electric only in
the N--S sector \cite{wit}. On the other hand the group
 which has an electric action both on
the N--S and R--R sector is $Sl(8,R)$.
 This follows from the alternative decomposition of the
${\bf 56}$ \cite{cj}, \cite{hull}:
\begin{equation}
{\bf 56} \, {\stackrel{Sl(8,R) }{\longrightarrow}} \,
  {\bf 28}
  \oplus \, {\bf 28}
\end{equation}
We can look at the intersection of the ST--duality
 group with the maximal electric group:
\begin{equation}
SL(2,\IR)\otimes O(6,6) \, \cap \, Sl(8,\IR) \,
= \, Sl(2,R)\, \otimes\, Sl(6,\IR) \,
\otimes \, O(1,1).
\end{equation}
Consideration of this subgroup allows to split
 into N--S and R--R parts the maximal
electric solvable algebra. Let us define it.
 Let ${Solv}\left (E_{7(7)}/SU(8)\right )$
be the complete solvable algebra.
The electric part is defined by:
\begin{equation}
{Solv}_{el} \, \equiv \, {Solv}\left (E_{7(7)}/SU(8)\right )
 \, \cap \, Sl(8,\IR) \, =
\,{Solv}\left (Sl(8,\IR)/O(8)\right )
\label{sl8so8}
\end{equation}
Hence we have that:
\begin{equation}
\mbox{dim}_{\bf R} \, {Solv}_{el} \, = 35
\end{equation}
One immediately verifies that the non--compact coset
manifold $Sl(8,\IR)/O(8)$ has
maximal rank, namely $r=7$, and therefore the electric
solvable algebra has once more
the standard form as in eq.\ref{simstru} where ${\cal H}$
is the Cartan subalgebra
of $Sl(8,\IR)$, which is the same as the original Cartan
subalgebra of $E_{7(7)}$
and the sum on positive roots is now restricted to those
that belong to $Sl(8,R)$.
These are $28$. On the other hand the adjoint representation
 of $Sl(8,\IR)$ decomposes
under the $Sl(2,\IR)\otimes Sl(6,\IR) \otimes O(1,1)$
  as follows
\begin{equation}
{\bf 63} ~~~ {\stackrel{Sl(2,\IR)\otimes Sl(6,\IR)
\otimes O(1,1)}{\longrightarrow}}
\, \left ({\bf 3},{\bf 1} ,{\bf 1} \right ) \, \oplus \,
\left ({\bf 1},{\bf 35} ,{\bf 1} \right ) \, \oplus \,
\left ({\bf 1},{\bf 1} ,{\bf 1} \right )
\, \oplus \, \left ({\bf 2},{\bf 6} ,{\bf 2} \right )
\label{qui}
\end{equation}
Therefore the N--S generators of the electric solvable
 algebra are the $7$ Cartan generators
plus the $16=1 \oplus 15$ {\it positive roots} of
$Sl(2,\IR)\otimes Sl(6,\IR)$.
 The R--R generators
are instead the {\it positive weights} of the
$\left ({\bf 2},{\bf 6} ,{\bf 2} \right )$
representation. We can therefore conclude that:
\begin{equation}
\mbox{dim}_{\bf R} \, {Solv}_{el} \, = 35 \, = 12 \mbox{R--R} \,
\oplus  [(15+1)+7]\, \mbox{N--S}
\end{equation}
\par
Finally it is interesting to look for the maximal abelian
subalgebra
 of the electric solvable
algebra.
It can be verified that the dimension of this algebra is 16,
corresponding to 8 R--R and 8 N--S.

\section{Considerations on non--maximally extended supergravities}
Considerations similar to the above can be made for all
the  non maximally extended or matter coupled
supergravities for which the solvable Lie algebra is not of maximal rank.
 Indeed, in the present case, the set of positive roots entering in formula
(\ref{iwa}) is a proper subset of the positive roots
of $U$, namely those which are not orthogonal to the
whole set of roots defining the non--compact Cartan generators.
As an example, let us analyze the coset ${O(6,22) \over O(6)
\otimes O(22) }\otimes {Sl(2,\IR)\over U(1)}$
corresponding  to a $D=4$, $N=4$ supergravity theory obtained
compactifying type IIA string theory
 on $K_3 \times T_2$ \cite{dn}, \cite{sei}, \cite{ht}.
The product $Sl(2,\IR) \otimes O(6,22) $ is the U--duality group
of this theory, while the
  ST--duality group is
 $Sl(2,\IR) \otimes O(4,20 ) \otimes O(2,2)$. The latter acts on the
 moduli space of $K_3 \times T_2$ and on the dilaton--axion system.
Decomposing the U--duality group with respect to the ST duality group
 $Sl(2,\IR) \otimes O(4,20 ) \otimes O(2,2)$
we get:
\begin{eqnarray}
  \label{o622}
  \mbox{adj}(Sl(2,\IR) \otimes O(6,22)) &=& \mbox{adj}Sl(2,\IR) \nonumber\\
&+&  \mbox{adj}O(4,20)+\mbox{adj}O(2,2)+
({\bf 1},{\bf 24},{\bf 4})
\end{eqnarray}
The R--R fields belong to the subset of positive roots
of U contributing to $\IG_s$ which are also positive
 weights of the ST--duality group,
namely in this case those defining the $({\bf 1},{\bf 24},{\bf 4})$
representation. This gives us 48 R--R fields.
The N--S fields, on the other hand, are selected by taking those
positive roots of U entering
 the definition of $\IG_s$, which are also positive roots of ST, plus those
 corresponding to the non--compact generators ($\cH_k$) of the
 $U$--Cartan subalgebra.
\par
In our case we have:
\begin{eqnarray}
  &&\mbox{dim} U =\mbox{dim}O(6,22) + \mbox{dim}Sl(2,\IR) = 381 \nonumber\\
&&  \mbox{\# of positive roots of U }= 183 \nonumber\\
&&  \mbox{\# of positive roots of U not contributing to $\IG_s$ }
\nonumber\\
&& \qquad \qquad\quad = 183 -(\mbox{dim}U/H - \mbox{rank}U/H) =56
\nonumber\\
&& \mbox{dim} ST =  \mbox{dim}O(4,20) +  \mbox{dim}O(2,2) +
\mbox{dim}Sl(2,\IR) = 285
\nonumber\\
&&  \mbox{\# of positive roots of ST } = {1\over 2} (285 -15) = 135
\nonumber\\
&&   \mbox{\# of positive roots of ST  contributing to $\IG_s$ }= 135-56=79
\nonumber\\
&&   \mbox{\# of N--S }= 79 + \mbox{rank}U/H = 79+7 = 86 \nonumber\\
&& \mbox{dim} (U/H) =\mbox{dim}\IG_s = 48+86 = 134.
\end{eqnarray}
The maximal abelian ideal $\cA$ of $\IG_s$ has dimension
 64 of which 24 correspond to R--R fields while 40 to N--S fields.
\par
In an analogous way one can compute the number of N--S and R--R
fields for other
non maximally extended supergravity theories.
\section{Conclusions}
In this note we used a particular parametrization of non compact
coset spaces
underlying various duality symmetries in terms of solvable Lie algebras.
In this way we found a natural splitting between R--R and N--S scalars.
For maximal supergravities the associated cosets, and therefore
the solvable algebras, have
maximal rank while this is not the case for non maximal and/or
matter coupled supergravities.
\par
The generators of the maximal abelian ideal of solvable Lie algebras
correspond
to the Peccei--Quinn symmetries of the theory.
\par
Part of them pertain to the R--R scalars and part to the N--S scalars.
Contrary to naive reasoning R--R scalars do not always correspond to
translational symmetries.
This can be traced back to Chern--Simons couplings in the original theory.
\par
Partial supersymmetry breaking with vanishing cosmological constant
 appears also to be related to the gauging of nilpotent generators
 of the solvable Lie algebra.
\par
It is hoped that some of the aspects of solvable Lie algebra discussed
in this paper may unreveal
some nonperturbative properties underlying superstring dynamics.
\section*{Acknowledgments}
Two of us (P. F. and S. F.)  would like to thank
K. Stelle and M. Gunaydin respectively for stimulating discussions.

\end{document}